\documentclass{article}
\usepackage{spconf,amsmath,graphicx,hyperref}
\usepackage{amssymb}
\usepackage{booktabs}
\usepackage{subcaption}
\usepackage{xcolor}

\title{One-Step Voice Conversion by Learning kNN Transport in WavLM Space}

\twoauthors
 {Anton Selitskiy}
	{Stony Brook University\\Electrical and Computer Engineering\\Stony Brook, NY 11790 USA}
 {David Millard}
	{University of Rochester\\Electrical and Computer Engineering\\Rochester, NY 14627 USA}

\begin{document}
%
\maketitle
\begin{abstract}
Voice conversion (VC) systems fall into two families: non-parametric embedding-space methods, which need no trained model but degrade on short target utterances, and spectrogram-based neural architectures, which achieve strong quality via multi-module pipelines with tens of millions of parameters. We propose $k$NN-FM-VC, a single conditional flow-matching network that learns to approximate the $k$NN-VC mapping between WavLM embedding distributions of source and target speakers, replacing explicit pointwise $k$NN matching with a neural regressor trained on $k$NN-generated pairs. The model is conditioned on the target speaker via cross-attention and FiLM, and trained under three Gaussian conditional paths (Schr\"odinger bridge, straight line, and constant-variance Gaussian tube), enabling few-step sampling. Unlike Phoneme Hallucinator, which uses an upsampling stage followed by kNN matching, our 13M-parameter model performs conversion with a single learned network and supports one-step inference. On LibriSpeech, the one-step Gaussian Bridge achieves lower WER and higher estimated speech quality than FreeVC and Phoneme Hallucinator. Relative to $k$NN and $k$DOT, it substantially reduces WER. 
\end{abstract}

\begin{keywords}
Voice conversion, Flow matching, Optimal transport, Schr\"{o}dinger bridge
\end{keywords}

\section{Introduction}
\label{sec:intro}

Voice conversion (VC) transforms speech from a source speaker while preserving linguistic content and transferring the characteristics of a target speaker. Recent self-supervised speech representations provide a useful latent space for this task, separating conversion from waveform generation and enabling direct manipulation of high-level speech features. A central challenge, however, is learning a reliable transformation between source and target embedding distributions, particularly when only limited target-speaker data are available. This motivates methods that can learn the structure of this transport directly rather than relying on explicit frame-level matching at inference. Qualitative audio samples are available on our project page.\footnote{\url{https://anton-selitskiy.github.io/kNN-VC-FM/}}

\subsection{Spectrogram-based Approaches}
Traditional VC systems often operate on spectrogram or mel-spectrogram representations and use speaker or \(F_0\) conditioning. FreeVC~\cite{freevc} uses a VITS-based CVAE-GAN with a WavLM content bottleneck and a normalizing flow conditioned on a pretrained speaker embedding. MeanVC~\cite{meanvc} uses streaming ASR features, timbre and speaker encoders, and a DiT decoder to generate converted mel-spectrograms. We exclude MeanVC because it is trained on Mandarin speech.

\begin{figure}[t]
    \centering
    \includegraphics[width=0.95\linewidth]{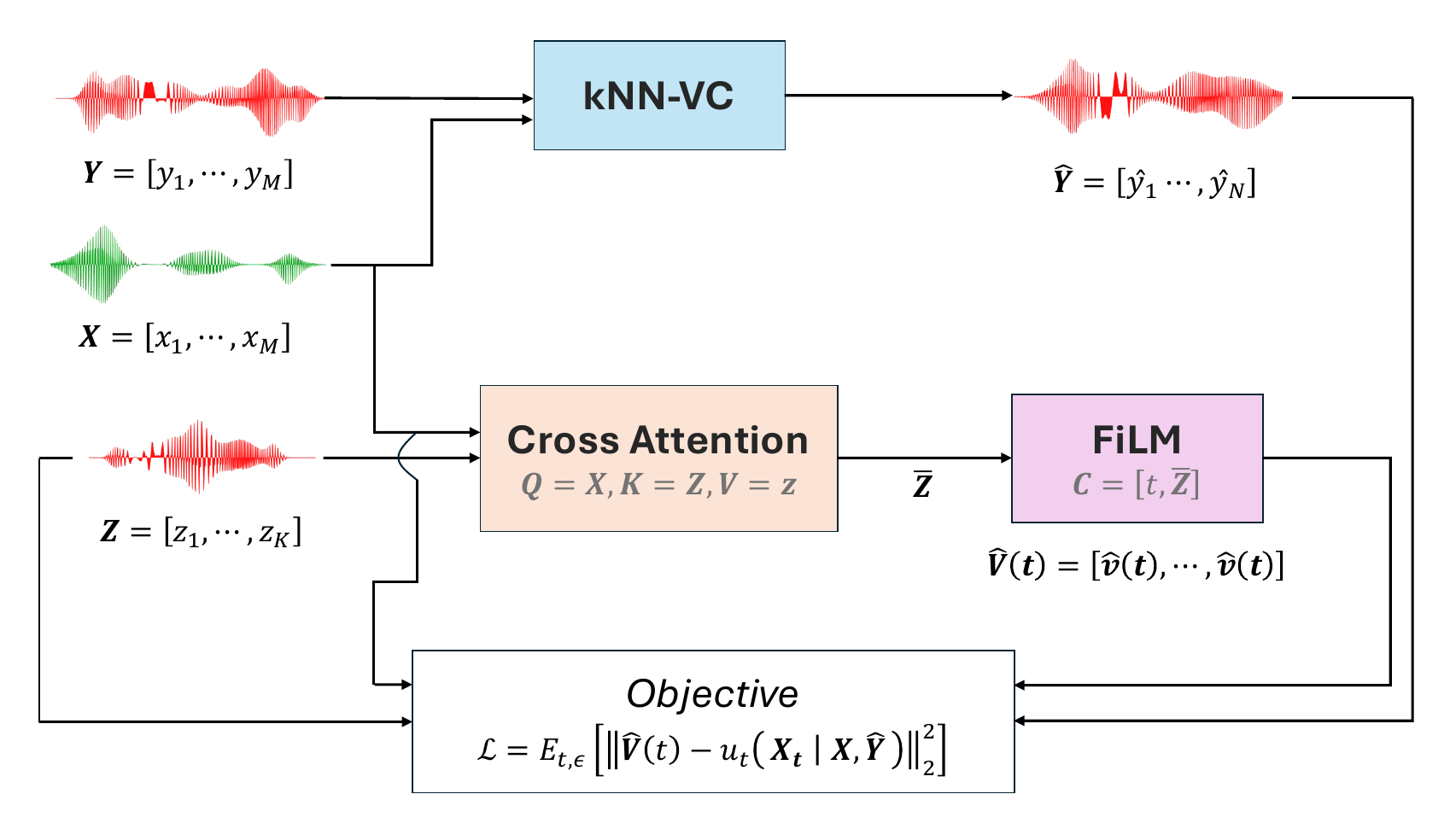}
    \caption{An overview of the core flow matching method. Flow matching transports source WavLM embeddings toward $k$NN-derived target embeddings while conditioning on target-speaker information.}
    \label{fig:overview}
\end{figure}

\subsection{Embedding-space, Non-parametric Approaches}
Embedding-space methods instead perform conversion directly on pretrained self-supervised features. $k$NN-VC~\cite{VCNN} extracts 1024-dimensional \textit{WavLM Large}~\cite{wavlm} representations,
$\mathbf{x}=[x_1,\ldots,x_M]$ and $\mathbf{y}=[y_1,\ldots,y_N]$,
and maps each source embedding to target embeddings using $k$NN regression before waveform reconstruction with HiFi-GAN. $k$DOT-VC~\cite{DVCOT} retains this training-free framework but replaces $k$NN averaging with discrete optimal transport. Both methods degrade with short target utterances as the target embedding set becomes sparse. MKL-VC~\cite{got} improves speech clarity under this setting but weakens speaker preservation. Phoneme Hallucinator~\cite{shan2024phoneme} instead expands the target embedding set with a learned generative model before applying $k$NN matching, resulting in a two-stage system with  $\sim$19M parameters.

\subsection{Proposed Method}
We replace explicit $k$NN matching with a conditional flow-matching (CFM) network that learns the source-to-target mapping in WavLM space. The model uses cross-attention and FiLM for target-speaker conditioning, eliminating both pointwise nearest-neighbor search and a separate up-sampling stage. Our $\sim$13M-parameter model is about $30\%$ smaller than the most similar model, Phoneme Hallucinator, and even supports single-step inference.

To our knowledge, CFM has not previously been used to learn the conversion map directly in WavLM embedding space between $k$NN-generated pairs, which is the approach we take here. Unconditional flow-matching for one-to-one voice conversion was used in~\cite{VCOT}. Our contributions are (i) We introduce a conditional flow-matching model that learns $k$NN-based transport directly in WavLM embedding space; (ii) we condition the model on the target speaker using cross-attention and FiLM, and study three Gaussian probability paths; (iii) we present a principled analysis of extending the OT paradigm to the task of VC; (iv) we substantially improve intelligibility over discrete $k$NN/$k$DOT baselines while supporting single-step inference; and (v) generalizes to other embedding-space VC methods.

\section{Voice Conversion Algorithm}
\label{sec:2}

\begin{table*}[t]
\centering
\small
\setlength{\tabcolsep}{3.5pt}
\begin{tabular}{llccccccc}
\toprule
\textbf{Model} & \textbf{Steps} & \textbf{WER $(\downarrow)$}
& \textbf{UTMOSv2 $(\uparrow)$}
& \textbf{FAD$_s$}
& \textbf{FAD$_t$}
& \textbf{$\Delta\textbf{FAD}_\text{rel}$ $(\uparrow)$}
& \textbf{Source SIM $(\downarrow)$}
& \textbf{Target SIM $(\uparrow)$} \\
\midrule
Schr\"odinger Bridge
& 1  & $23,[22,24]$ & $3.04 \pm 0.42$ & $0.85$ & $1.03$ & $21.2\%$ & $0.55,[0.54,0.56]$ & $\mathbf{0.30},[0.29,0.31]$ \\
& 5  & $23,[22,24]$ & $3.11 \pm 0.43$ & $0.78$ & $0.96$ & $23.1\%$ & $0.64,[0.63,0.65]$ & $0.28,[0.27,0.28]$ \\
& 50 & $23,[22,24]$ & $3.11 \pm 0.42$ & $0.75$ & $0.93$ & $\mathbf{24.0\%}$ & $0.65,[0.64,0.66]$ & $0.27,[0.26,0.27]$ \\
\addlinespace

Gaussian Bridge
& 1  & $23,[23,24]$ & $\mathbf{3.12 \pm 0.37}$ & $0.93$ & $1.07$ & $15.1\%$ & $0.55,[0.54,0.56]$ & $0.29,[0.28,0.30]$ \\
& 5  & $23,[22,24]$ & $3.11 \pm 0.42$ & $0.81$ & $0.98$ & $21.0\%$ & $0.64,[0.63,0.65]$ & $0.27,[0.26,0.28]$ \\
& 50 & $23,[22,24]$ & $3.10 \pm 0.43$ & $0.78$ & $0.95$ & $21.8\%$ & $0.65,[0.65,0.66]$ & $0.26,[0.25,0.27]$ \\
\addlinespace

Gaussian Tube
& 1  & $23,[22,24]$ & $3.08 \pm 0.38$ & $0.91$ & $1.04$ & $14.3\%$ & $0.55,[0.54,0.56]$ & $0.29,[0.29,0.30]$ \\
& 5  & $23,[22,24]$ & $3.09 \pm 0.40$ & $0.80$ & $0.96$ & $20.0\%$ & $0.64,[0.63,0.65]$ & $0.27,[0.26,0.27]$ \\
& 50 & $23,[22,24]$ & $\mathbf{3.12 \pm 0.42}$ & $0.75$ & $0.92$ & $22.7\%$ & $0.66,[0.65,0.66]$ & $0.26,[0.25,0.26]$ \\
\midrule

Discrete
& kNN  & $50,[47,53]$ & $2.50 \pm 0.51$ & $1.22$ & $1.41$ & $\underline{15.6\%}$ & $0.18,[0.17,0.18]$ & $0.48,[0.47,0.49]$ \\
& kDOT & $48,[46,50]$ & $2.51 \pm 0.47$ & $1.12$ & $1.29$ & $15.2\%$ & $0.17,[0.16,0.17]$ & $\underline{0.51},[0.50,0.51]$ \\
& MKL  & $72,[61,82]$ & $2.36 \pm 0.64$ & $3.51$ & $3.70$ & $5.4\%$ & $0.28,[0.27,0.30]$ & $0.27,[0.26,0.28]$ \\
\addlinespace

FreeVC
& -- & $\underline{24},[23,26]$ & $\underline{2.86 \pm 0.39}$ & $2.58$ & $2.80$ & $8.5\%$ & $0.31,[0.30,0.31]$ & $0.28,[0.27,0.29]$ \\

Phoneme Hallucinator
& -- & $25,[24,26]$ & $\underline{2.86 \pm 0.43}$ & $1.65$ & $1.79$ & $8.5\%$ & $0.34,[0.33,0.34]$ & $0.36,[0.35,0.37]$ \\
\bottomrule
\end{tabular}

\caption{Voice conversion performance for different probability paths and baselines. WER and SIM are reported as the mean with a 95\% confidence interval in brackets. FAD$_s$ and FAD$_t$ denote FAD relative to the source and target distributions, respectively, and $\Delta\mathrm{FAD}_{\text{rel}}=(\mathrm{FAD}_t-\mathrm{FAD}_s)/\mathrm{FAD}_s$. Source SIM measures similarity to the source speaker, while target SIM measures similarity to the target speaker. UTMOSv2 is reported as the mean $\pm$ sample standard deviation.}
\label{tab:comparison}
\end{table*}

\subsection{Flow-Matching}

We first introduce the idea of \textit{Lagrangian flows}. Assume that a sample $x$ is transformed at time $t=1$ into a sample $y$ along a trajectory $\xi(x,t)$ satisfying the dynamical system
\begin{equation}
\begin{cases}
    \dot{\xi}(x,t) = v(\xi(x,t),\, t),\\
    \xi(x,0) = x.
\end{cases}
\end{equation}
The short notation is often used $\xi(x,t) \equiv \xi_t(x) \equiv x_t$, with $x_0 = x.$  If points $x$ are distributed at $t=0$ according to $\mu_0 = \mu$, the push-forward measure $\mu_t = \xi_{t\sharp}\mu$ satisfies the continuity equation $\partial_t\mu_t - \nabla\cdot(v\mu_t) = 0$.

\subsubsection{Conditional Probability Paths} Following the Flow Matching (FM) framework of~\cite{lipman2023flow} and its generalization~\cite{tong2024improving}, we construct $v$ by regressing a neural network $v_\theta$ against a per-example (conditional) vector field. For a Gaussian conditional probability path
\begin{equation}
p_t(x \mid x_0, x_1) = \mathcal{N}\!\left(x \,\middle|\, m_t(x_0,x_1),\, \sigma_t(x_0,x_1)^2 I\right),
\end{equation}
the unique vector field generating this path is~(Theorem 3 of~\cite{lipman2023flow} applied to $(x_0,x_1)$ instead of $x_1$; see also~\cite{tong2024improving})
\begin{equation}
\label{eq:general_ut}
u_t(x \mid x_0, x_1) = \frac{\sigma_t'}{\sigma_t}\big(x - m_t\big) + m_t',
\end{equation}
and the conditional flow-matching (CFM) objective
\begin{equation}
\mathcal{L}_{\mathrm{FM}}(\theta) = \mathbb{E}_{t,\, q,\, p_t(\cdot\mid x_0,x_1)}\left\| v_\theta(x,t) - u_t(x\mid x_0,x_1)\right\|^2
\end{equation}
has gradients identical to the (intractable) marginal FM objective~\cite{lipman2023flow,tong2024improving}. $q(x_0,x_1)$ is the joint source-target distribution. 

\subsubsection{Application to VC}  In our setting $x_0=\mathbf{x}$ and the $k$NN target $x_1=\hat{\mathbf{y}}^{\text{kNN}}$ are the endpoints of the conditional path, and we additionally condition the network on an auxiliary target-speaker recording $\mathbf{z}$ (with $\mathbf{z}=\mathbf{y}$ at inference). We use the linear mean
\begin{equation}
m_t = (1-t)x_0 + tx_1,
\end{equation}
and connect the endpoints with a Gaussian bridge $m_t + \sigma_t\boldsymbol{\epsilon}$, $\boldsymbol{\epsilon}\sim\mathcal{N}(0,I)$. Since $(\mathbf{x}, \hat{\mathbf{y}}^{\text{kNN}})$ is only an approximate coupling (not the true OT plan between the marginal distributions), the choice of variance schedule $\sigma_t$ is not uniquely determined by the endpoints alone; we compare three such schedules in Sec.~\ref{sec:variants}.

The model is conditioned on the target speaker in two complementary ways. First, we apply cross-attention between the source sequence $\mathbf{x}$ (queries) and the auxiliary sequence $\mathbf{z}$ (keys and values), aligning source content with target speaker characteristics at the frame level. Second, we inject a global speaker representation $\bar{\mathbf{z}}$ together with time $t$ through FiLM layers with a learnable scale $\gamma$ and shift $\beta$:
\begin{equation}
\mathrm{FiLM}(h) = \gamma(\bar{\mathbf{z}}, t) \odot h + \beta(\bar{\mathbf{z}}, t).
\end{equation}

The flow-matching objective, applying Eq.~\eqref{eq:general_ut} to our conditional path,  (outlined in Figure~\ref{fig:overview})
\begin{equation}
\mathcal{L}_{\mathrm{FM}} = \mathbb{E}_{t,\boldsymbol{\epsilon}} \left\| v_\theta(x_t, t, \mathbf{z}) - u_t(x_t \mid x_0, x_1) \right\|^2,
\end{equation}
where $u_t$ takes the specific form given for each variant in Sec.~\ref{sec:variants} below. We add a speaker consistency loss
\begin{equation}
\mathcal{L}_{\mathrm{SC}} = 1 - \cos\!\big(\bar{\mathbf{z}}, \bar{\mathbf{y}}^{\text{pred}}\big),
\end{equation}
where, instead of full integration, $\mathbf{y}^{\text{pred}}$ is obtained by a one-step approximation
$
\mathbf{y}^{\text{pred}} \approx x_t + (1 - t)\, v_\theta(x_t, t, \mathbf{z}).
$
The total training loss is therefore:
\begin{equation}\label{eq:loss}
\mathcal{L} = \mathcal{L}_{\mathrm{FM}} + 0.1\, \mathcal{L}_{\mathrm{SC}}.
\end{equation}

\subsubsection{Inference}
At inference, the target utterance provides the conditioning sequence, so we set $\mathbf{z}=\mathbf{y}$. Starting from the source embeddings $x_0=\mathbf{x}$, we numerically integrate
\begin{equation}
    \frac{dx_t}{dt}
    = v_\theta(x_t,t,\mathbf{z}),
    \qquad
    x_0=\mathbf{x},
\end{equation}
from $t=0$ to $t=1$ using the desired number of integration steps. The endpoint $x_1=\hat{\mathbf{y}}$ is the converted WavLM representation, which is decoded to waveform $\hat{y}$ using the HiFi-GAN vocoder.

\subsection{Variants of FM objectives}
\label{sec:variants}

Each variant below is a particular choice of $\sigma_t$  in Eq.~\eqref{eq:general_ut}.

\subsubsection{Schr\"odinger Bridge}

Following the entropic-OT construction of SB-CFM~\cite{tong2024improving}, we take a Brownian-bridge variance schedule
\begin{equation}\label{eq:3:BB}
    \sigma_t = \sigma\sqrt{t(1-t)},
\end{equation}
which vanishes at both endpoints and is generated by the vector field
\begin{equation}
    u_t(x_t \mid x_0, x_1) = (x_1 - x_0) + \frac{1-2t}{2t(1-t)}\left(x_t - m_t\right).
\end{equation}
As $\sigma \to 0$ this reduces to the straight-line vector field below; for $\sigma>0$ it corresponds to the simulation-free approximation of the Schr\"odinger bridge between the (empirical) source and pseudo-target distributions~\cite{tong2024improving}.

\subsubsection{Gaussian Bridge with Straight Line}

We use the Brownian-bridge variance (\ref{eq:3:BB})
and regress the network onto the constant drift
\begin{equation}
    u_t(x_t \mid x_0, x_1) = x_1 - x_0,
\end{equation}
omitting the score-correction term $\frac{1-2t}{2t(1-t)}(x_t-m_t)$ that appears in the exact generating field for this path. When $(x_0,x_1)$ pairs are drawn from an approximately optimal-transport coupling, the resulting displacement fields can exhibit reduced ambiguity and straighter marginal trajectories~\cite[Th.~4.2]{flowm}. Motivated by this behavior, we retain the SB noise schedule $\sigma_t$ as a time-dependent perturbation while regressing onto the coupled endpoint displacement $x_1-x_0$, rather than the exact Gaussian-path vector field.

\subsubsection{Gaussian Tube}

As a third variant, we use a constant variance that is fixed at the value the Schr\"odinger-bridge schedule attains at its midpoint, $\sigma_t \equiv \sigma/\sqrt{2}$, again giving
\begin{equation}
    u_t(x_t \mid x_0, x_1) = x_1 - x_0.
\end{equation}
Unlike the straight-line variant, this schedule does not shrink to zero at $t=0,1$, so samples are not exactly concentrated at the endpoints; we found this to give a useful trade-off between path straightness and robustness to the fact that $(x_0,x_1)$ are only approximate transport pairs. Constant-tube paths appear in stochastic-interpolant work, but this construction appears to be new to VC/FM.

\section{Experimental Setup and Results}

\subsection{Training and Evaluation Data} 
We train  our flow-matching network entirely on  LibriSpeech train-clean-100~\cite{panayotov2015librispeech}  and evaluate all models on corresponding \textit{test} set.

\subsection{Implementation Details}
Audio is represented by 1024-dimensional features from layer~6 of a frozen \textit{WavLM Large} encoder. Pairs are constructed by pooling frames from up to eight source and ten target utterances and averaging the $k=4$ cosine-nearest target frames for each source frame. Speaker conditioning uses a separate target utterance cropped to 200 frames, with $\bar{\mathbf z}$ obtained by masked temporal averaging. The $13.26$M-parameter network uses four-head cross-attention at width 1024, 128-dimensional time and speaker embeddings, and four FiLM residual blocks. We train with batch size 64 using AdamW ($\mathrm{lr}=10^{-4}$, weight decay $10^{-4}$), mixed precision, and unit gradient clipping. We sample $t\sim\mathrm{Beta}(2,2)$, use $\sigma=0.03$, and optimize velocity MSE plus a speaker-consistency loss weighted by $0.1$, detailed in Eq.~(\ref{eq:loss}).

\subsection{Metrics} We report WER using Whisper~\cite{whisper} with JiWER alignment, UTMOSv2~\cite{utmosv2}, and FAD using pre-activation VGGish embeddings~\cite{fadtk}. Speaker similarity (SIM) follows the SEED-TTS evaluation protocol~\cite{meanvc}, using a WavLM-Large/ECAPA speaker verifier, where same-speaker pairs yield scores of approximately $0.7$.

\subsection{Comparison with Baselines} 
Table~\ref{tab:comparison} shows that the proposed FM variants achieve substantially lower WER and higher estimated speech quality than the discrete $k$NN, $k$DOT, and MKL baselines. Relative to FreeVC, the one-step FM models improve intelligibility, estimated quality, and target-relative FAD while achieving comparable SIM. Compared with Phoneme Hallucinator, the FM variants improve WER and estimated quality, and achieve a substantially larger relative FAD separation between source and target (14–24\% vs. 8.5\%), but obtain lower target-speaker SIM. Notably, $k$NN and $k$DOT achieve the highest SIM despite substantially worse WER and UTMOSv2, indicating that speaker-verification similarity does not necessarily track intelligibility or perceptual quality and should be interpreted alongside the other metrics. Table~\ref{tab:pairwise} complements the aggregate results by showing how these differences are distributed across individual conversions. For WER, most pairs are unchanged, but improvements occur more often than degradations against both learned baselines. SIM is nearly balanced against FreeVC, whereas Phoneme Hallucinator more frequently achieves higher similarity than Gaussian Bridge. Thus, the pairwise analysis reinforces the primary advantage of Gaussian Bridge in intelligibility while exposing a trade-off with speaker-verification similarity against Phoneme Hallucinator.

\begin{table}[t]
\centering
\small
\setlength{\tabcolsep}{4pt}
\begin{tabular}{lccccc}
\toprule
\textbf{Baseline}
& \multicolumn{3}{c}{\textbf{WER (\%)}}
& \multicolumn{2}{c}{\textbf{SIM (\%)}} \\
\cmidrule(lr){2-4}
\cmidrule(lr){5-6}
& \textbf{Better} & \textbf{Tie} & \textbf{Worse}
& \textbf{Better} & \textbf{Worse} \\
\midrule
FreeVC
& $\mathbf{27.1}$ & $57.6$ & $15.3$
& $\mathbf{53.4}$ & $46.6$ \\
Phoneme Hallucinator
& $\mathbf{34.5}$ & $52.4$ & $13.1$
& $25.2$ & $\mathbf{74.8}$ \\
\bottomrule
\end{tabular}
\caption{Pairwise outcomes of the one-step Gaussian Bridge relative to each baseline over 1,310 conversions. Better and worse are defined from the perspective of Gaussian Bridge: lower is better for WER, while higher is better for SIM.}
\label{tab:pairwise}
\end{table}

\subsection{Comparison of FM Variants} 
At 50 integration steps, a Holm-corrected Wilcoxon signed-rank test found no significant WER differences among the three probability paths. Schr\"odinger Bridge achieves slightly higher SIM, and also the largest relative FAD separation. Thus, no path consistently dominates; each trades off speaker similarity, FAD, and efficiency.

\subsection{\textbf{Ablation}: Sensitivity to Integration Steps} 
We further compare each variant across 50, 5, and 1 integration steps. WER remains similar across step counts, while speaker-verification SIM is highest at one step for all three probability paths and decreases with additional integration. In contrast, relative FAD separation increases with more integration steps for all three variants, while changes in UTMOSv2 remain small. These results indicate a trade-off rather than a uniform advantage for fewer steps: one-step inference maximizes SIM and matches WER/UTMOSv2 at a fraction of the cost, but additional integration steps improve the model's distributional separation from the source domain.

\section{Discussion}

Learning transport in WavLM space primarily improves intelligibility and quality over discrete matching, while path choice and NFE trade off SIM, FAD, and efficiency. Notably, $k$NN and $k$DOT obtain higher SIM despite substantially worse WER and UTMOSv2, suggesting SIM captures speaker-discriminative cues that do not track perceptual quality and should be interpreted jointly with other metrics. Source-reference SIM reveals a further asymmetry: our FM variants retain more source-speaker character than some baselines; at one step, however, target-speaker SIM matches or exceeds FreeVC and MKL, and all variants attain the highest relative FAD separation overall, suggesting SIM and FAD capture complementary rather than redundant aspects of conversion quality. These findings position single-step FM as an efficient alternative to explicit embedding-space matching.


\section{Conclusion}

We introduce a conditional flow-matching framework for voice conversion that learns $k$NN-derived transport directly in WavLM space with target-speaker conditioning, under three Gaussian probability paths. The proposed method matches or exceeds substantially larger baselines in intelligibility and quality while using a smaller, single-step model.

\vfill\pagebreak

\section{Compliance with Ethical Standards}
This study used the publicly available LibriSpeech corpus under its CC BY 4.0 license. No new data were collected and no interaction or intervention with human participants was conducted; therefore, no additional ethical approval was required.

\section{Acknowledgment}

The first author was supported in part by the New York State Center of Excellence in Data Science under Grant C25089A007. The authors have no relevant financial or nonfinancial interests to disclose. The first author thanks Arip Asadulaev, Rostislav Korst, and Alexander Tikhonov for discussions of paper~\cite{VCOT} and of the preliminary results. Both authors thank Mark Bocko and Andrea Cogliati for their continued interest in this work.

\bibliographystyle{IEEEbib}
\bibliography{refs}

\end{document}